\documentclass[twocolumn]{raa}
\usepackage{graphicx,times}
\usepackage{natbib}
\usepackage{amssymb,amsmath}
\bibpunct{(}{)}{;}{a}{}{,}

\usepackage[pagebackref=true]{hyperref}

\hypersetup{colorlinks = true, linkcolor = green, anchorcolor = red, citecolor = blue, filecolor = red, pagecolor = red, urlcolor = red}

\begin{document}

\title{The halo concentration and mass relation traced by satellite galaxies}

\volnopage{ {\bf 20XX} Vol.\ {\bf X} No. {\bf XX}, 000--000}
   \setcounter{page}{1}

\author{Qing Gu\inst{1,2}, Qi Guo\inst{1,2}, Tianchi Zhang\inst{3}, Wenting Wang\inst{4,5}, Quan Guo\inst{6}, Liang Gao\inst{1,2,7}
      }
\institute{ Key Laboratory for Computational Astrophysics, National Astronomical Observatories, Chinese Academy of Sciences, Beijing 100101, China; {\it guoqi@nao.cas.cn}\\
         \and
         School of Astronomy and Space Science, University of Chinese Academy of Sciences, Beijing 10049, China\\
         \and
         Beijing Planetarium, Beijing Academy of Science and Technology, Beijing 100044, China\\
	     \and
         Department of Astronomy, Shanghai Jiao Tong University, Shanghai 200240, China\\
         \and 
         Shanghai Key Laboratory for Particle Physics and Cosmology, Shanghai 200240, China\\
         \and 
         Shanghai Astronomical Observatory, Chinese Academy of Sciences, 80 Nandan Road, Shanghai 200030, China\\
         \and
         Institute of Computational Cosmology, Department of Physics, University of Durham, Science Laboratories, South Road, Durham DH1 3LE, UK\\
 \vs \no
   {\small Received 20XX Month Day; accepted 20XX Month Day}
}

\abstract{We study the relation between halo concentration and mass ($c-M$ relation) using the Seventh and Eighth Data Release of the Sloan Digital Sky Survey (SDSS DR7 and DR8) galaxy catalogue. Assuming that the satellite galaxies follow the distribution of dark matter, we derive the halo concentration by fitting the satellite radial profile with a Nararro Frank and White (NFW) format. The derived $c-M$ relation covers a wide halo mass range from $10^{11.6}$ to $10^{14.1}\rm \ M_{\odot}$. We confirm the anti-correlation between the halo mass and concentration as predicted in cosmological simulations. Our results are in good agreement with those derived using galaxy dynamics and gravitational lensing for halos of $10^{11.6}-\ 10^{12.9}$\ M$_{\odot}$, while they are slightly lower for halos of  $10^{12.9}-\ 10^{14.1}$\ M$_{\odot}$. It is because blue satellite galaxies are less concentrated, especially in the inner regions. Instead of using all satellite galaxies, red satellites could be better tracers of the underlying dark matter distribution in galaxy groups.
}

\authorrunning{Gu et al.}            
   \titlerunning{The halo concentration and mass relation}  
   \maketitle

\section{Introduction} 
\label{sec:intro}

In the cold dark matter (CDM) paradigm of structure formation, one of the most fundamental aspects to investigate is the formation and evolution of dark matter halos. The density profiles of dark matter halos determine the potential and also reflect the formation histories. It has been discovered that the density profiles of dark halos follow a uniform functional form, the NFW profile, regardless of differences in halo properties \citep{Navarro1996}. The NFW density profile is presented as a function of the radial distance $r$ with two free parameters,
\begin{equation}
    \rho(r)=\frac{\rho_0}{\frac{r}{r_{\rm s}}(1+\frac{r}{r_{\rm s}})^2},
    \label{NFW}
\end{equation}
where $r_{\rm s}$ is a scale radius that separates the internal and  external regions, with $\rho \propto r^{-1}$ in the innermost regions of the halo and $\rho \propto r^{-3}$ in the outer regions of the halo. $\rho_{0}$ is four times the density at $r_{\rm s}$.

The halo concentration characterizes the flatness of the density profile, which is defined as $c = r_{\rm vir}/r_{\rm s}$. Here $r_{\rm vir}$ is the virial radius of a given halo. There exists a relation between the halo concentration and mass in N-body simulations\citep{Navarro1997,Bullock2001,Eke2001}. Low mass halos have higher concentrations, while high mass halos have lower concentrations. A recent study \citep{Wang2020} extended this relation down to halo masses as low as $10^{-3}\ \rm M_{\odot}$. \cite{Bullock2001} found that the concentration scales linearly with the cosmic scale factor at fixed halo mass ($c \propto (1+z)^{-1}$). However, later works found a more complex mass and redshift dependence. The evolution of the halo concentration depends on the mass, i.e. low mass halos evolve more strongly with redshift than high mass halos \citep{Wechsler2002,Zhao2003,Correa2015}. At the very massive end, the concentration is almost independent of halo mass and only evolves very slightly with redshift \citep{Zhao2003,Zhao2009,Gao2008}. Furthermore, the density profile of the dark matter halos can be influenced by baryonic processes, as found in hydro-dynamical simulations \citep{Schaller2015,Butsky2016}. For example, active galactic nuclei (AGN) feedback could flatten the DM density profile and reduce the concentration \citep{Duffy2010,Martizzi2013,Suto2017}.

In observations, the halo profiles are usually measured using gravitational lensing and X-ray data,  \citep[e.g.][]{Merten2015, Mantz2016}. However, the halo concentration inferred from lensing observations could be overestimated due to the projection effects \citep{Meneghetti2007} or the presence of massive background structures \citep{Coe2012}. \citet{Martinsson2013} found a large scatter around the $c-M$ relation using galaxy dynamics in the DiskMass survey \citep{Bershady2010}. Most of the observed $c-M$ relations are measured in rather narrow mass ranges, either focusing on galaxy-size halos or on cluster-size halos. There are no $c-M$ relations covering both the galaxy-size halos and the cluster-size halos simultaneously. 

Another tracer of the underlying dark matter density profile is the population of satellite galaxies. It has been demonstrated that the number density profile of satellites is an unbiased tracer of the total mass distribution in rich clusters \citep{Carlberg1997,Diemand2004,Gao2004,Wang2018} and the stellar mass density profile of satellites tend to trace the distribution of total mass in galaxy groups spanning a wide mass range \citep{Wang2021}. 
Using the SDSS DR7 and DR8 galaxy sample, \citet{Wang2014} (hereafter W14) and \cite{Guo2012} (hereafter G12) measured the radial distribution of satellite galaxies in halos of $\sim 10^{11.6}-10^{14.1}\ \rm M_{\odot}$, a much wider mass range than those in most previous works. Based on these studies, we revisit the $c-M$ relation between galaxy-size halos and cluster-size halos. 

\section{Data}

We use the satellite profiles measured by W14 and G12 to calculate the halo concentrations. W14 selects primary galaxies from the spectroscopic catalogue of the New York University Value-Added Galaxy Catalogue (NYU-VAGC) \citep{Blanton2005} based on the Seventh Data Release of the Sloan Digital Sky Survey (SDSS DR7; \citet{Abazajian2009}). The primary candidates are requested to have apparent (Petrosian) $r$-band magnitude brighter than $m_{r}$ = 16.6. They further restrict the primary galaxies to be isolated so that: each must (i) be at least one magnitude brighter than any companion within a projected radius of $r_{\rm p}$ = 0.5 Mpc and within a line-of-sight velocity difference $c|\Delta z| <$ 1000 km $\rm s^{-1}$, and (ii) be the brightest object within $r_{\rm p} <$ 1 Mpc and $c|\Delta z| <$ 1000 km $\rm s^{-1}$. Satellite galaxies are identified in the SDSS DR8 photometric catalogue \citep{Aihara2011} and corrected for background contamination statistically. Briefly, around each primary galaxy, they select all photometric galaxies with $r$-band apparent model magnitude brighter than $m_{r}$ = 21 and with projected distance $r_{\rm p} <$ 0.5 Mpc. Then they compute galaxy counts as a function of projected separation, $r_{\rm p}$, $r$-band apparent magnitude, $m_{r}$, and color, $g-r$. For each $r_{\rm p}$ bin, they subtract the expected number of galaxies based on the average number of background galaxies in each $(m_{r},\ g-r)$ bin. The excess number with respect to a homogeneous background is assumed to be the number of satellite galaxies around the given primary galaxy.

G12 used slightly different selections. They select primary galaxies from the spectroscopic catalogue with $m_{r} \leq 17.77$ and satellite galaxies from photometric catalogue with $m_{r} \leq 20.5$ in SDSS DR8. The isolation criteria are also slightly different from W14 that they request no neighboring galaxies brighter than $M_{\rm p} + 0.5$, where $M_{\rm p}$ is the absolute magnitude of primary galaxy, within a projected radius of 2$R_{\rm inner}$, and with redshift difference of $|z_{\rm c}-z_{\rm s}^{\rm neigh}| < 0.002$ if the neighbour galaxy has a spectroscopic redshift, or $|z_{\rm c} - z_{\rm p}^{\rm neigh}| < 2.5 \sigma_{\rm p}^\ast$ if only photometric redshift is available, where $\sigma_{\rm p}^\ast$ is the measurement error of the photometric redshift. The satellite number density profiles are estimated by accounting for the excess number of galaxies within the projected radius $R_{\rm inner}$ with respect to the averaged galaxy number densities between $R_{\rm inner}$ and $R_{\rm outer}$. In G12, the primaries are divided into three luminosity bins with absolute magnitudes of (-21.75, -20.75], (-22.5, -21.5] and (-23.5,-22.5], respectively. Their corresponding ($R_{\rm inner}$, $R_{\rm outer}$) are (1.25, 2.5), (1.08, 2.16) and (1.06, 1.73) $r_{\rm 200}$. Here $r_{\rm 200}$ is the virial radius, within which the average enclosed density is 200 times the critical density ($\rho_{\rm crit}$) of the Universe.  
 
To obtain the $c-M$ relation, we need the corresponding halo masses in different stellar mass or luminosity bins of galaxies. W14 adopted the mean halo mass of primary galaxies in each stellar mass bin using the semi-analytic galaxy catalogues based on the Millennium and Millennium-II simulations \citep{Guo2011}. G12 used the mean halo mass of primary galaxies in each luminosity bin according to the $M_* - M_{\rm 200}$ relation predicted by the abundance matching method \citep{Guo2010}. Here $M_{\rm 200}$ is the virial mass and $M_{\rm 200} = \frac{4}{3}\pi200\rho_{\rm crit} r_{200}^3$.

Integrating the three-dimensional NFW density profile (Eq. \ref{NFW}) along the line-of-sight direction one can obtain the projected surface density profile of dark halo as a function of projected radius $r_{\rm p}$,
\begin{equation}
\Sigma(r_{\rm p})=2\int_{0}^{\infty}{\rho(r_{\rm p}, z)}dz,
\label{NFWp}
\end{equation}
This integral can be solved analytically \citep{Bartelmann1996} as,

\begin{equation}
\Sigma(x)=\left\{
\begin{array}{ll}
\frac{2\rho_0 r_{\rm s}}{x^2-1}\left[ 1-\frac{2}{\sqrt{(1-x^2)}}\rm arctanh\sqrt{\frac{1-x}{1+x}}\right] &x<1,\\
\frac{2\rho_0 r_{\rm s}}{3} &x=1,\\
\frac{2\rho_0r_{\rm s}}{x^2-1}\left[ 1-\frac{2}{\sqrt{(x^2-1)}}\rm arctan\sqrt{\frac{x-1}{1+x}}\right] &x>1.
\end{array}
\right.
\end{equation}
where $x=r_{\rm p}/r_{\rm s}$\footnote{For G12, they applied background subtraction process to the projected NFW profile. It is denoted as $\hat{\Sigma}(r_{\rm p})=\Sigma(r_{\rm p})-\frac{2}{3r_{200}^2}\int_{r_{200}}^{2r_{200}}r_{\rm p}\Sigma(r_{\rm p})dr_{\rm p}$}.
 By fitting to the observed surface density profiles, one could estimate the parameters $\rho_0$ and ${r_{\rm s}}$ and derive the halo concentration defined as 
\begin{equation}
    c_{200} = \frac{r_{200}}{r_{\rm s}}.
\end{equation}

\begin{figure}
\centering
\includegraphics[width=\columnwidth]{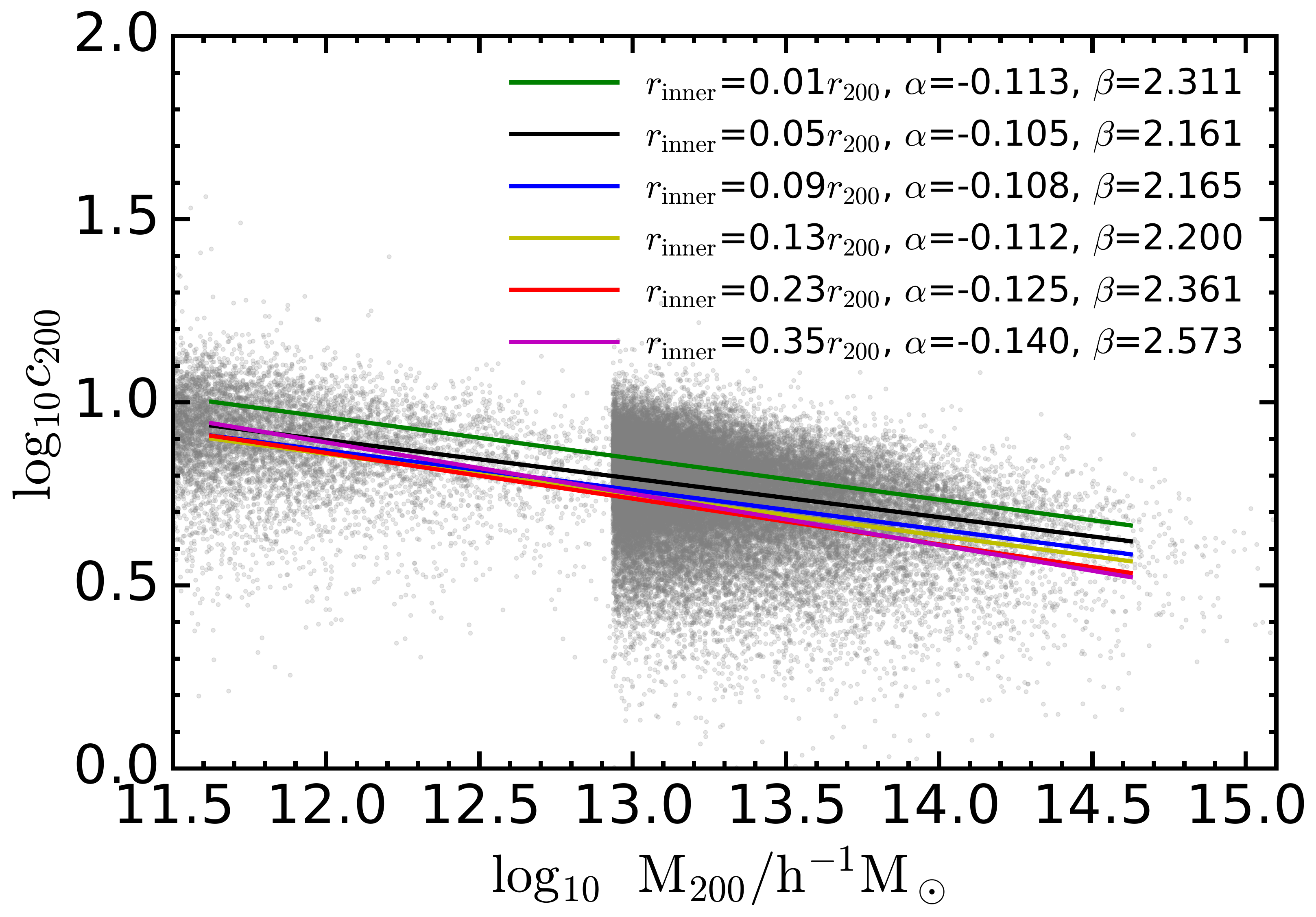}
    \caption{The relations of halo concentration and mass for halos in the MS and MSII. The dark matter distribution in each halo is fitted with a NFW format from an inner radius $r_{inner} $ to the virial radius $r_{\rm 200}$. The grey points show the halo concentration with the inner radius of 0.05$r_{\rm 200}$. The two clouds correspond to results from the  MSII (higher resolution and smaller volume) at low masses and the MS (lower resolution and larger volume) at high masses. The median value of the $c-M$ relation is fitted using eq. \ref{scale}, denoted with the black line. Lines with different colours correspond to different choices of the inner radii as indicated in the top right corner, along with the corresponding fitting parameters.}
    \label{fig:scale}
\end{figure}

It has been noticed that different choices of the inner radius would lead to different fitting parameters \citep[e.g.][]{Neto2007}. W14 and G12 adopted different inner radii to avoid the deblending effects, we quantify its effect on the derived concentration parameters by varying the inner radius cut in cosmological simulations. We split halos at the mass of $10^{13}\ \rm M_{\odot}$. For massive halos, we use the Millennium Simulation \citep[MS,][]{Springel2005}, while for less massive ones, we use the Millennium-II Simulation \citep[MSII,][]{Boylan-Kolchin2009}. Since results are less sensitive to the outer radii, we fix it to $r_{\rm 200}$ all through this analysis. We first fit the dark matter profiles with the inner radius of 0.05$r_{\rm 200}$ as presented with grey dots in Fig. \ref{fig:scale}. The median value of $c_{200}$ as a function of halo mass is fitted with a power law format (black line),
\begin{equation}
     {\rm log}  c_{200}=\beta + \alpha {\rm log} M_{200} /h^{-1}\rm M_\odot
\label{scale}
\end{equation}
 where $\alpha = -0.105$ and $\beta = 2.161$. Then we present the fitted $c-M$ relations with lines of different colours corresponding to different choices of inner radii. It shows that the differences in the derived halo concentration could be up to 38.5\%. To remove the variation due to the choice of different inner radii, we scale all the concentrations to that with $r_{\rm inner}$ = 0.05$r_{\rm 200}$ as follows,  
\begin{equation}
    c_{200}^{0.05}=c_{200}^x10^{\beta_{0.05}-\beta_x}(M_{200}/h^{-1}\rm M_\odot)^{\alpha_{0.05}-\alpha_x}
\end{equation}
where ${x}$ denote the inner radius in unit of $r_{\rm 200}$.

\section{Results}

\begin{figure*}
	\includegraphics[width=1\textwidth]{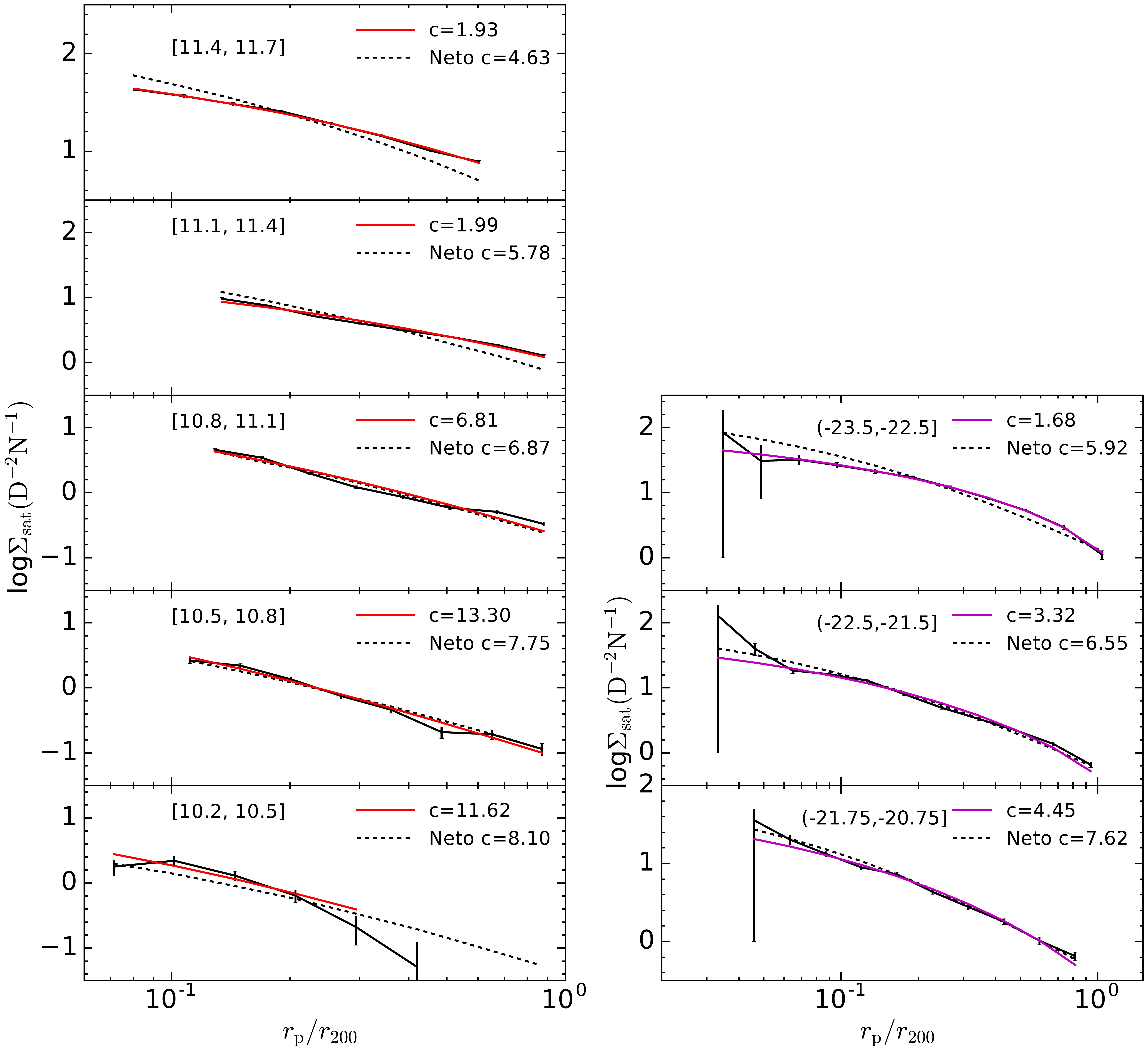}
	
    \caption{Projected number density profiles for satellite galaxies in groups and clusters. The left panel shows the distribution of satellites brighter than $r$-band apparent magnitude $m_{r}$ = 21 and for primaries in different stellar mass bins in W14. The logarithm of the stellar mass ranges (log($M_\ast/\rm M_{\odot}$)) are indicated by the quoted numbers in the brackets. The right panel shows the results in G12 for satellites with an absolute magnitude brighter than -19 and around primaries in different luminosity bins indicated by the quoted numbers in the brackets. The $y$-axis is the satellite numbers per primary and per unit surface area. The measurements are shown by the solid black lines with error bars generated by bootstrapping technique. The red and magenta solid lines show the fitting profiles assuming a projected NFW format. The concentration values in legends have been scaled to concentrations with $r_{\rm inner}$ = 0.05$r_{\rm 200}$. The black dashed lines denote the expected dark matter profiles with typical concentration  \citep{Neto2007} at halo masses either derived from the semi-analytical galaxy catalogue \citep{Guo2011} (W14) or from the abundance matching method  \citep{Guo2010} (G12).}
    \label{fit}
\end{figure*}

Fig.~\ref{fit} shows the stacked projected number density profiles of satellite galaxies ($\Sigma_{\rm sat}$) as a function of projected separation $r_{\rm p}/r_{\rm 200}$. The left panel shows the results based on W14 and the right panel shows the results based on G12. The black solid curves are the measurements in observation and the dashed curves are the expected DM density profiles assuming a typical $c-M$ relation by \citet{Neto2007}. The error bars are 1-$\sigma$ dispersions among 100 and 1000 bootstrapped subsamples for W14 and G12, respectively. 

We fit a projected NFW profile to the projected density profile (see eq. \ref{NFWp}). The best fits are obtained by minimizing the $\chi^2$ statistic, between the logarithm of $\Sigma_i$ and the projected NFW profile,
\begin{equation}
\chi^2 = \sum_{i=1}^{N_{\rm bins}}[\frac{ \rm log_{10} \Sigma_{\it i}-\rm  log_{10} \Sigma_{NFW}(\rho_0;\it r_{\rm s})}{\sigma_i}]^2,
\end{equation}
where $\sigma_i$ represents the error in the $i$-th bin. 
The best fits are presented with red and magenta curves in Fig. \ref{fit} and the corresponding concentration is indicated in the upper right corner of each panel. For the lowest mass bin of W14, we remove the last data point as its error is too large\footnote{The few outermost data points in this bin are quite noisy and decrease very fast. This is because W14 adopted very strict isolation criteria to select primary galaxies. All primaries tend to be in regions with very low density. The global background for subtraction could be higher than the true local background around the primaries, introducing slight over-subtraction in the background number counts. This is not a problem for massive primaries, as the true satellite counts are significantly above the background level. However, the over-subtraction becomes prominent for the outskirts of the lowest mass bin, where the background is significantly dominating.}. 

It shows that the NFW profile leads to good fits. Our estimated concentration parameters are consistent with those predicted in simulations \citep{Neto2007} for relatively low mass and intermediate mass systems, while for massive primaries, our estimation is significantly lower.

\begin{figure*}
\centering
\includegraphics[width=0.8\textwidth]{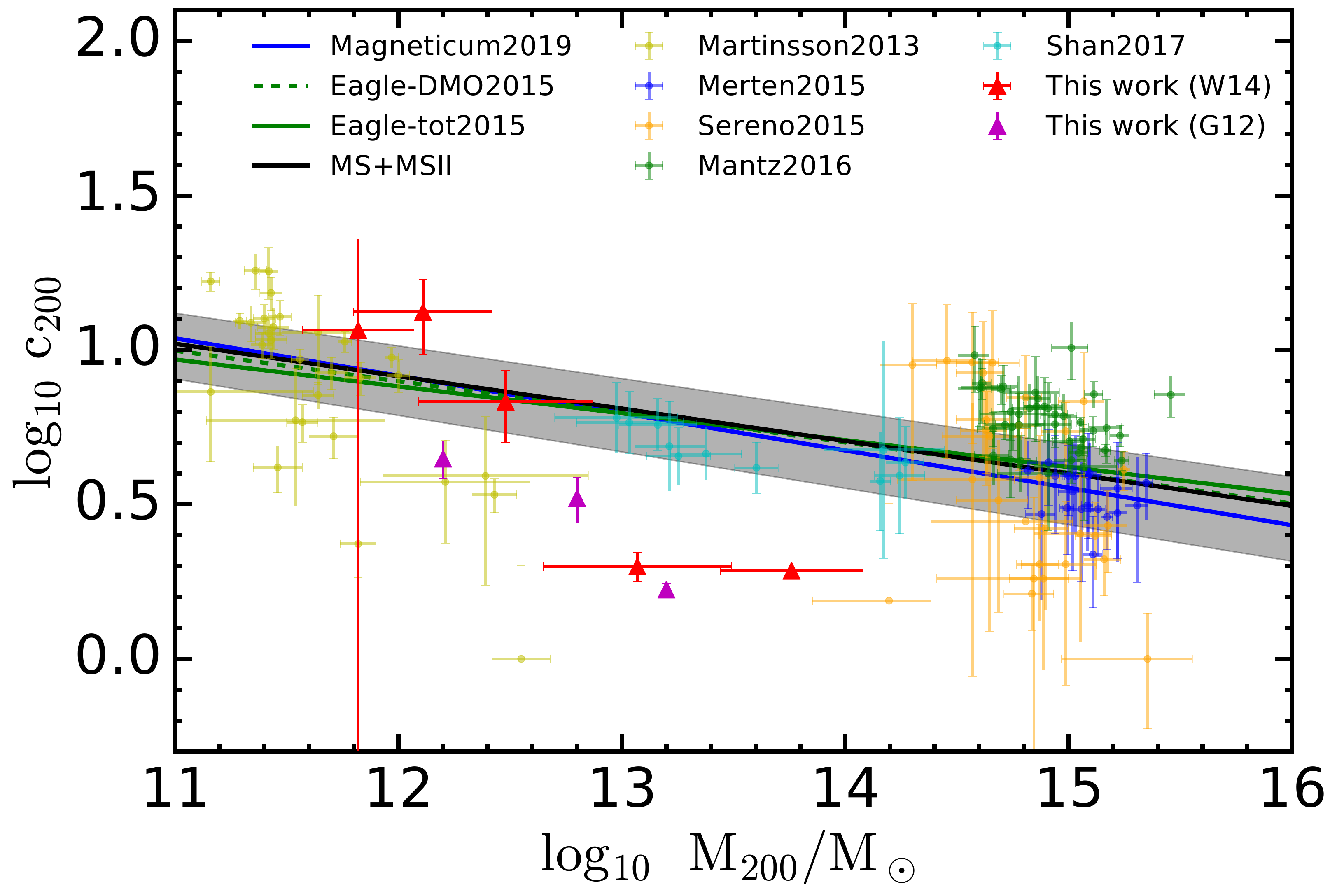}

    \caption{The concentration-mass relation. The red and magenta triangles with error bars show our results based on W14 and G12, respectively. Different colour points show other measurements in the literature. The curves represent the relations in different simulations. The black solid line and grey shade region show the median and the 68 percentile scatter of the $c-M$ relation in the MS and MSII.}
    \label{fig:cm}
\end{figure*}

  The main results are presented in Fig. \ref{fig:cm}. Halo concentrations inferred by satellite distributions show a clear declination with halo masses, from $10^{11.6}\ \rm M_{\odot}$ to $10^{14.1}\ \rm M_{\odot}$. The trend is consistent with those found previously in simulations and in observations. 
  
 Measurements in the literature using different methods are compiled in Fig. \ref{fig:cm}. Yellow crosses are the low mass systems ($<10^{13}\ \rm M_{\odot}$) estimated using the dynamics of 30 spiral galaxies in the DiskMass survey \citep{Martinsson2013}. Weak lensing measurements at $10^{13}-10^{14}\ \rm M_{\odot}$ are taken from the Canada–France–Hawaii Telescope Stripe 82 Survey (CS82), the redMaPPer cluster catalogue and the LOWZ/CMASS galaxy sample of the Sloan Digital Sky Survey-III Baryon Oscillation Spectroscopic Survey Tenth Data Release \citep{Shan2017}. The results are estimated at $0.2<z<0.6$ and then scaled to $z = 0$, assuming the redshift evolution from \citet{Klypin2016}. At higher masses, the $c-M$ relations are estimated using the X-ray and lensing measurements \citep[][]{Merten2015,Mantz2016,Sereno2015}. 

\begin{figure*}
\centering

\includegraphics[width=0.49\textwidth]{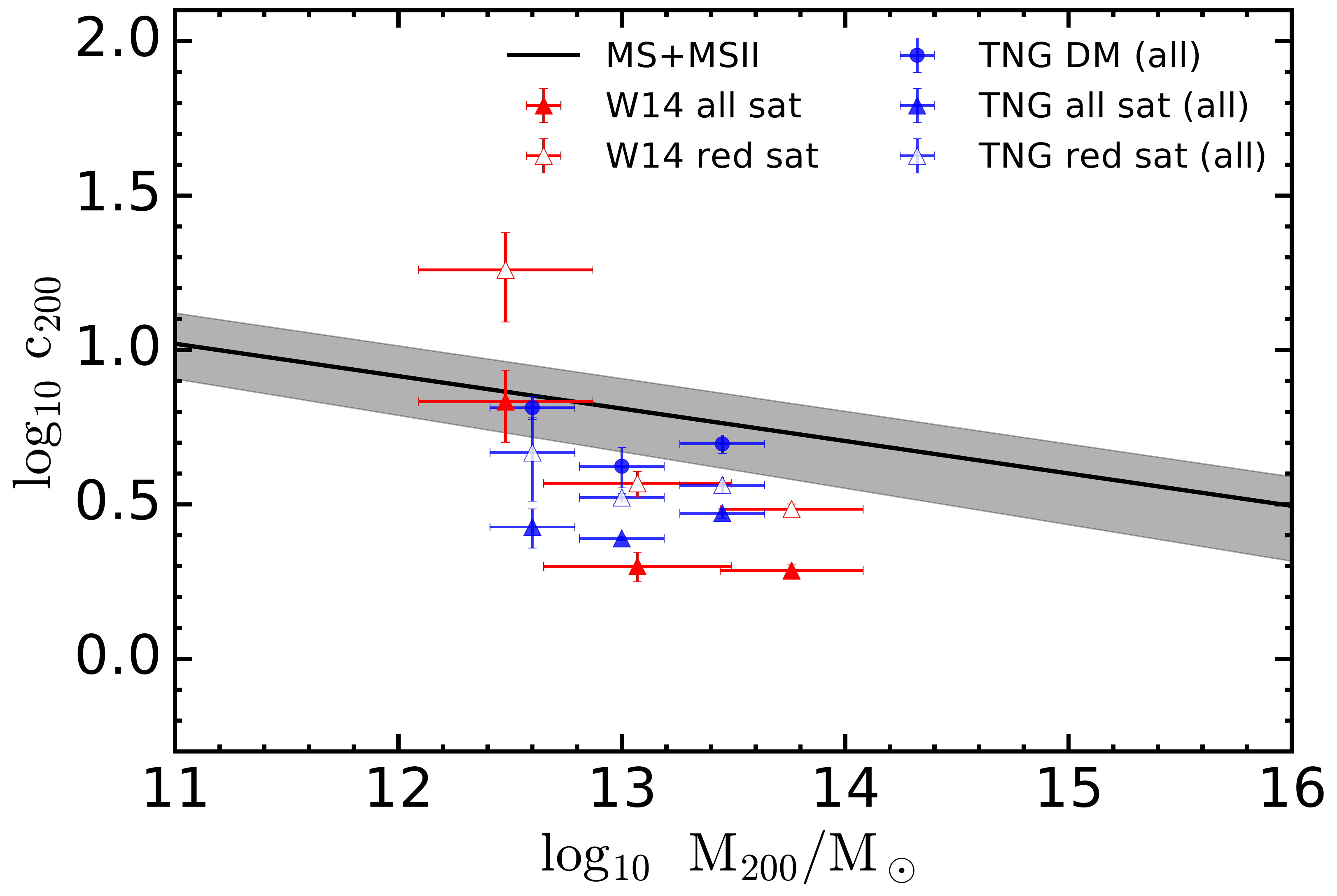}
\includegraphics[width=0.49\textwidth]{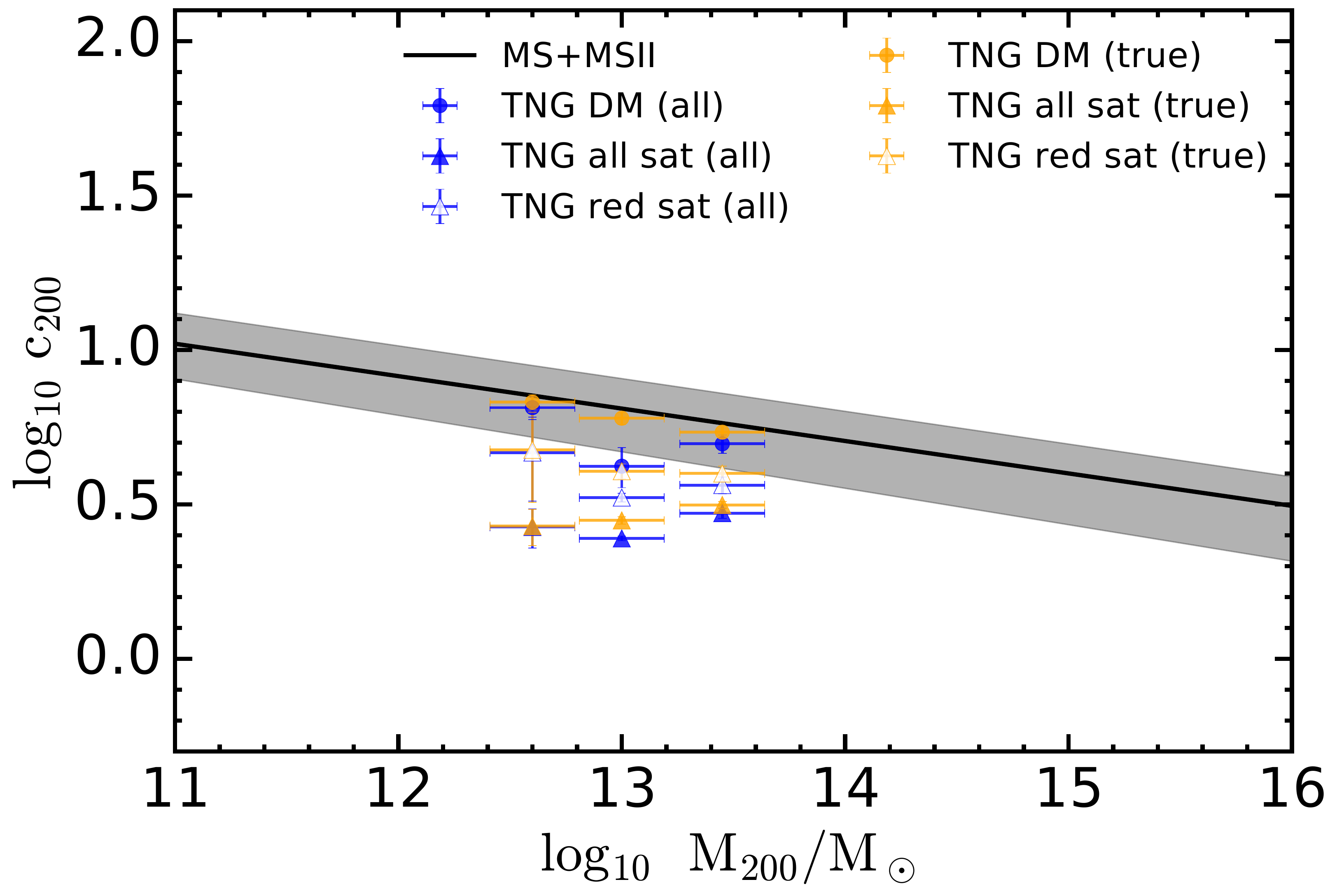}
    \caption{Comparison of the concentration-mass relations based on all satellites (filled triangles) and based on red satellites (empty triangles). The black solid line and grey shade region are the same as in Fig. \ref{fig:cm}. \textit{Left panel}: The red and blue triangles show the results based on W14 and all primaries in IllustrisTNG simulations, respectively. The blue dots show the concentration measured from the actual dark matter distribution in IllustrisTNG. \textit{Right panel}: The blue dots and triangles are the same as in the left panel. The orange results show the $c-M$ relations for primaries that are true central galaxies in IllustrisTNG simulations.}
    \label{fig:cm_TNG}
\end{figure*}
The theoretical predictions of the $c-M$ relations using N-body and hydro-dynamical cosmological simulations are also included for comparison. The black solid line shows the result of the MS and MSII and the grey shade region shows the 68 percentile scatter. The blue solid line shows the $c-M$ relation in Magneticum hydrodynamic simulation \citep{Ragagnin2019}. The two green lines show the $c-M$ relations from the dark matter-only (DMO) and the full Evolution and Assembly of Galaxies and their Environment (EAGLE) simulations \citep{Schaller2015}. All simulations predict a power law relation of the $c-M$ relation, consistent with each other and with the estimated $c-M$ in the literature using galaxy dynamics, weak lensing and X-ray, though the latter has a large scatter. Note that the EAGLE-tot and EAGLE-DMO show similar $c-M$ relations, suggesting the baryonic processes merely change the density profile in the mass range considered here. 

It shows that the simulation predictions and the measurement using galaxy dynamics, weak lensing and X-ray are broadly consistent with each other. Our estimated halo concentrations follow the predicted $c-M$ relation at halo masses below $10^{12.9}\ \rm M_\odot$, while at intermediate masses, $[10^{12.9},\ 10^{14.1}]\ \rm M_\odot$ results based on W14 are relatively lower. This leads to a steeper slope of the $c-M$ relation when fitting across the full mass range $[10^{11.6},\ 10^{14.1}]\ \rm M_\odot$. Results based on G12 have even lower halo concentrations. This could be due to the selection of relatively brighter satellite galaxies in G12. As demonstrated in G12, brighter galaxies tend to be less concentrated.

To better understand how satellite galaxies could trace the dark matter distribution in halos, we present the $c-M$ relation estimated using satellite galaxies in the IllustrisTNG simulations \citep{Nelson2018,Nelson2019} in Fig. \ref{fig:cm_TNG}. We generate mock catalogues by assigning each TNG galaxy a redshift based on its line-of-sight distance and peculiar velocity assuming an observer at the origin of the coordinates in the TNG100-1 and TNG300-1 boxes at $z=0$. We use the same stellar mass ranges, magnitude limit and isolation criteria for selecting primaries as those for W14, but just use real satellites which belong to the same friends-of-friends (FOF) group as the primary and are more massive than $10^{8.2}\ \rm M_\odot$. We select primary galaxies with stellar mass less than $10^{11.1}\ \rm M_\odot$ from the TNG100-1 simulation and more massive primary galaxies from TNG300-1. For each primary, we use the surrounding DM particles that belong to the same FOF group to calculate the density profile of the host DM halo. The comparison of $c-M$ relations based on W14 and IllustrisTNG simulations is shown in the left panel of Fig. \ref{fig:cm_TNG}. Our selected sample of primary galaxies in TNG has a small satellite contamination fraction up to 10.5\%.

It shows that TNG satellite galaxies tend to be less concentrated than dark matter, which is consistent with the results of \citet{McDonough2022}. Satellite galaxies could thus be a biased tracer of the underlying matter distribution. Moreover, it is known that red satellite galaxies and blue satellite galaxies fall into clusters/groups at different time and have different spatial distributions. Thus we investigate the halo concentrations traced by red and all satellite galaxies separately in Fig. \ref{fig:cm_TNG}. We select red satellites in TNG based on $g-r$ colour cuts of 0.64, 0.58 and 0.52 for satellites with stellar mass larger than 10$^{10.2}\ \rm M_\odot$, between 10$^{9.2}\ \rm M_\odot$ and 10$^{10.2}\ \rm M_\odot$ and between 10$^{8.2}\ \rm M_\odot$ and 10$^{9.2}\ \rm M_\odot$, respectively. For each stellar mass bin, the colour cut corresponds to the trough between the blue and red peaks of the galaxy colour distribution in the TNG100-1. For observation, the projected number density profiles of red satellites are directly taken from W14. It shows that red galaxies have higher concentrations than that of all satellite galaxies, and their concentrations are closer to those traced by dark matter at halo masses $[10^{12.9},\ 10^{14.1}]\ \rm M_\odot$. Compared to full types of satellite galaxies, red satellite galaxies are better tracers of the matter distribution in galaxy groups. This is true in both real observation and TNG. 

We notice that the concentrations of TNG satellite galaxies are slightly higher than those derived from SDSS at $\log_{10}M_{200}/\rm M_\odot>13$, this could be due to the  misidentification of the central galaxies in SDSS, i.e., observationally, the sample of isolated primary galaxies include a small fraction of contamination by satellite galaxies, and although we select primary galaxies in TNG following exactly the same criteria as W14, it is possible that the fraction of satellite contamination is different between SDSS and TNG primaries. To test the contamination effect, we compare the $c-M$ relations by using only those primaries which are real central galaxies in TNG simulations. The results are shown in the right panel of Fig. \ref{fig:cm_TNG}. The orange results show $c-M$ relations of the primaries which are real central galaxies and are located at the center of the host halo. The concentrations are slightly larger than those of all primaries as shown by the blue results. It demonstrates that satellite contamination can result in slightly smaller concentrations, but the difference is very small. It is thus likely that the real SDSS primaries may have a slightly larger fraction of satellite contamination compared with TNG primary galaxies at $\log_{10}M_{200}/\rm M_\odot>13$. \citet{Skibba2011} found that the fraction $f_{\rm BNC}$ (for ‘Brightest-Not-Central’) increases from $\sim$ 0.25 in low mass haloes (10$^{12}\ h^{-1} \rm M_\odot$ $\leq M \lesssim 2\times10^{13}\ h^{-1} \rm M_\odot$) to $\sim$ 0.4 in massive haloes ($M \gtrsim 5 \times 10^{13}\ h^{-1} \rm M_\odot$) in SDSS. The values are larger than our maximum fraction of contamination of 0.11 in TNG. Moreover, we note that possible offsets between the central coordinates of galaxies defined through optical photometry and the actual potential minimum positions might also result in smaller concentrations \citep{Neto2007}. In the real Universe, isolated primary galaxies do not necessarily reside exactly in the potential minimum, even if they are true central galaxies of the host dark matter halos, especially when the systems are not fully relaxed. This effect might not be realistically reflected in TNG.

\section{Conclusions}

The relation between halo concentration and mass is one of the fundamental relations in cosmology. Previous works on the $c-M$ relation usually focus on rather narrow mass ranges. Taking advantage of the measurement of the radial distribution of satellite galaxies in the SDSS groups and clusters by \citet{Wang2014} and \citet{Guo2012}, we are able to obtain the $c-M$ relation over a much wider mass range. 

We find an anti-correlation between the halo concentration and halo mass from galactic halos ($10^{11.6}\ \rm M_\odot$) to galaxy clusters ($10^{14.1}\ \rm M_\odot$). The trend is consistent with those reported in the literature. However, our halo concentration estimates at $[10^{12.9},\ 10^{14.1}]\ \rm M_\odot$ is lower than those estimated using weak lensing data and those found in simulations, leading to a stronger dependence of the halo concentration on halo masses. Similar results have been found by \citet{Collister2005} in the 2PIGG groups.

We find the population of red satellite galaxies trace better the distribution of dark matter than all satellites. The profiles of blue satellites are more flattened than those of red satellites \citep{Wang2014}. The deficit of blue satellite galaxies in the inner regions was also reported in 2PIGG groups \citep{Collister2005} and in SDSS BCG groups \citep{Budzynski2012}. Using the IllustrisTNG100 simulation, \citet{McDonough2020} found a similar result that red satellites are better tracers of the mass distribution in halos, whereas blue satellites show flattened profiles in inner regions, consistent with our results. Red satellites fell in the current host halos earlier. Their star
formations are quenched early and are more centrally concentrated due to dynamical frictions, which drag them towards the centers of their host dark matter halos. On the other hand, blue satellites fell in late and are so far affected less by dynamical frictions, which maintain their star formations and show more flattened radial distributions.

\begin{acknowledgements}
 This work is supported by the National Key Research and Development of China (No.2018YFA0404503), NSFC grants (No.12033008,11988101), the K.C.Wong Education Foundation, and the science research grants from the China Manned Space Project with NO.CMS-CSST-2021-A03 and NO.CMS-CSST-2021-A07. QG acknowledges support from the joint Sino-German DFG research Project “The Cosmic Web and its impact on galaxy formation and alignment” (DFG-LI 2015/5-1, NSFC No. 11861131006) and the support of the Shanghai International partners project (No.19590780200). WW acknowledges the support from NSFC (12022307) and the Yangyang Development Fund.
\end{acknowledgements}

\bibliography{ref}
\bibliographystyle{raa}

\end{document}